\documentclass[
  aps,
  pre,
  reprint,
  onecolumn,
  superscriptaddress,
  nofootinbib
]{revtex4-2}

\usepackage{amsmath}
\usepackage{amssymb}
\usepackage{bm}
\usepackage{graphicx}
\usepackage[hidelinks]{hyperref}

\newcommand{\PaperTitle}{
The Role of Odd Diffusivity in Multipoint Statistics of State-Dependent Observables
}

\newcommand{\FundingText}{
This research was supported by an appointment to the JRG Program at the APCTP through the Science and Technology Promotion Fund and Lottery Fund of the Korean Government. This was also supported by the Korean Local Governments - Gyeongsangbuk-do Province and Pohang City.
This work was supported by the National Research Foundation of Korea (NRF) grant funded by the Korea government (MSIT) (RS-2025-00557038).
}

\begin{document}

\title{\PaperTitle}

\author{Jong-Min Park}
\email{jongmin.park@apctp.org}

\affiliation{
Asia Pacific Center for Theoretical Physics,
Pohang 37673, Republic of Korea
}

\affiliation{
Department of Physics,
Pohang University of Science and Technology,
Pohang 37673, Republic of Korea
}

\begin{abstract}
Odd diffusivity is a transverse transport coefficient that appears when time-reversal and parity symmetries are broken. Its most distinctive signature is a probability flux perpendicular to a density gradient, generated by the antisymmetric part of the diffusion tensor. Here we show that for an arbitrary number of state-dependent observables measured at arbitrary times, their joint statistics are independent of the antisymmetric part of the diffusion tensor. Thus, the Lorentz flux does not contribute to any multipoint state-observable measurements. The result clarifies the separate roles of two effects that are often linked together by fluctuation--dissipation relations, i.e., odd mobility and odd diffusivity. This observation demonstrates that the anomalous correlations in odd-diffusive systems originate solely from odd mobility. It also allows the statistics of state-dependent observables in odd-diffusive systems to be computed using conventional Langevin dynamics without the antisymmetric diffusion part while preserving the antisymmetric mobility tensor. It implies that universal relations associated with state-dependent observables alone, such as nonlinear fluctuation--dissipation relations and generalized Green--Kubo relations, remain valid in odd-diffusive systems without modification. We verify our findings through mean back relaxation, the diffusion coefficient, the nonlinear fluctuation--dissipation relation, and a generalized Green--Kubo relation in diverse systems.
\end{abstract}

\maketitle

\section{Introduction}
Transport in systems with broken time-reversal and parity symmetry is often characterized by a transverse response.
The Hall effect is the canonical example: an antisymmetric component of the conductivity tensor generates a current perpendicular to an applied electric field.
Similar antisymmetric, or odd, transport coefficients have recently become central in active matter~\cite{vuijk2020lorentz,abdoli2021stochastic,muzzeddu2022active,wittmann2025confined} and chiral media~\cite{vega2022diffusive,hargus2025odd,hargus2025passive}, including odd viscosity~\cite{avron1998odd,banerjee2017odd,souslov2019topological,reichhardt2022active}, odd elasticity~\cite{scheibner2020odd,kole2021layered,yasuda2022time}, odd mobility~\cite{poggioli2023odd,ghimenti2023sampling,ghimenti2024transverse,ghimenti2024irreversible}, and odd diffusivity~\cite{hargus2021odd,langer2024dance,hargus2025flux}.

Odd diffusivity is the diffusive analogue of such Hall-like transport. It arises, for example, in the overdamped motion of Brownian particles under magnetic Lorentz forces~\cite{chun2018emergence,vuijk2019anomalous,abdoli2020nondiffusive,abdoli2020stationary,park2021thermodynamic}, in active chiral particles~\cite{hargus2021odd,muzzeddu2022active}, and in passive probes immersed in chiral active environments~\cite{hargus2025odd,hargus2025passive}. In two-dimensional odd-diffusive systems, the diffusion tensor contains an antisymmetric component. A density gradient then produces not only the usual diffusive flux parallel to the gradient, but also a perpendicular probability current, often called the Lorentz flux~\cite{abdoli2020stationary,kalz2022collisions}.

Several striking phenomena have been reported in odd-diffusive systems. Examples include superballistic spreading of overdamped Rouse dimers~\cite{shinde2022strongly}, enhancement of self-diffusion by repulsive collisions~\cite{kalz2022collisions, luigi2025self}, and oscillatory force autocorrelation functions in equilibrium odd-diffusive systems~\cite{kalz2024oscillatory}. They suggest that odd diffusivity can substantially affect time correlations.

However, some quantities are insensitive to the odd diffusivity.
In the Fokker--Planck equation with spatially uniform transport coefficients, the antisymmetric diffusion tensor contributes to the probability current but does not appear in the density evolution because the antisymmetric second-derivative operator vanishes~\cite{park2021thermodynamic}. Thus the one-time probability density is independent of the Lorentz flux. On the other hand, multipoint correlations are more subtle: they depend on the evolution of dynamics between different times and have played a central role in previous studies of anomalous odd-diffusive phenomena. It is therefore natural to ask whether the Lorentz flux, although invisible in one-time densities, can nevertheless affect time-correlation functions or more general multipoint statistics.

In this paper, we prove that the Lorentz flux does not affect a broad class of measurements. For arbitrary state-dependent observables evaluated at arbitrary times, and for any function of those observables, the expectation value is independent of the antisymmetric components of the diffusion tensor. This includes all multipoint correlation functions and the joint distributions of state observables. The proof is based on embedding the observables into an extended state space by introducing auxiliary variables associated with observables evaluated at arbitrary times.
We show that the antisymmetric part of the diffusion tensor
does not contribute to the evolution of the joint probability density in this extended state space.

The result gives a clear separation between the roles of odd diffusivity and odd mobility that are often linked by fluctuation--dissipation relations. Odd diffusivity generates a Lorentz flux in probability space, but this flux cannot be seen by coordinate-based multipoint statistics. Odd mobility, by contrast, generates transverse forces and nonreciprocal effective interactions~\cite{ghimenti2023sampling,ghimenti2024transverse,ghimenti2024irreversible,guo2025diffusion}. These forces are responsible for anomalous state-observable correlations such as collision-enhanced diffusion.

This result is practically important because it provides an alternative way to evaluate state-dependent observables. In general, odd-diffusive dynamics cannot be simulated or analyzed as an ordinary Langevin equation with white Gaussian noise~\cite{chun2018emergence,park2021thermodynamic}. An alternative approach is to introduce an auxiliary underdamped dynamics and take a singular small-mass limit. Such procedures may introduce finite-mass artifacts in numerical estimates. A direct criterion for when odd diffusivity can be ignored would therefore simplify both theory and simulation.
Consequently, all such correlations can be computed from ordinary white-noise Langevin dynamics in which the antisymmetric diffusion is set to zero while the antisymmetric mobility is retained.

Beyond its practical implications, this invariance also leads to important theoretical consequences. Any universal relation involving only state-dependent observables at multiple times remains valid in odd-diffusive systems whenever it is valid for ordinary Langevin dynamics. We demonstrate this for the nonlinear fluctuation--dissipation relation and for a generalized Green--Kubo relation. Conversely, the theorem identifies a fundamental limitation: detecting odd diffusivity requires observables that probe probability currents or other quantities beyond state-dependent multipoint statistics.

The remainder of this paper is organized as follows.
Section~\ref{sec:main} introduces the odd-diffusive dynamics and presents the main result.
Its derivation and physical consequences are discussed in Secs.~\ref{sec:derivation} and \ref{sec:consequence}, respectively.
Section~\ref{sec:examples} provides several numerical and analytical examples, including mean back relaxation, collision-enhanced diffusion, the nonlinear fluctuation--dissipation relation, and the generalized Green--Kubo relation.
Finally, Sec.~\ref{sec:conclusion} concludes the paper.

\section{Main result}
\label{sec:main}

We consider $N$ particles moving in two spatial dimensions with odd diffusivity.
The particle positions are collectively denoted by $\boldsymbol{x} = (\boldsymbol{x}_1, \boldsymbol{x}_2, \ldots, \boldsymbol{x}_N)^\mathrm{T}$, where $\boldsymbol{x}_n$ is the position of the $n$-th particle and the superscript $\mathrm{T}$ denotes the transpose.
All particles are assumed to be coupled to the same environment and therefore have identical mobility and diffusivity tensors, $\boldsymbol{\mu}$ and $\mathbf D$, respectively.
The time evolution of the probability density $P(\boldsymbol{x},t)$ is given by
\begin{equation}\label{eq:general_FP_eq}
    \dot{P} (\boldsymbol{x},t)
    = -\sum_n \boldsymbol{\nabla}^\mathrm{T}_{n}
    \left( \boldsymbol{\mu} \boldsymbol{f}_n(\boldsymbol{x},t) - \mathbf{D} \boldsymbol{\nabla}_n \right)
    P(\boldsymbol{x},t),
\end{equation}
where the overdot $\dot{\square}$ denotes a time derivative, $\boldsymbol{\nabla}_n$ is the gradient operator with respect to $\boldsymbol{x}_n$, and $\boldsymbol{f}_n (\boldsymbol{x},t)$ is the total force acting on the $n$-th particle.

The mobility and diffusion tensors are decomposed into their symmetric and antisymmetric parts as
\begin{equation}
    \boldsymbol{\mu} = \mu_0 \mathbf{I} + \kappa_\mu \boldsymbol{\epsilon},
    \qquad
    \mathbf{D} = D_0 \mathbf{I} + \kappa_D \boldsymbol{\epsilon},
\end{equation}
where $\mathbf{I}$ is the identity matrix, $\kappa_\mu$ and $\kappa_D$ characterize the strengths of the antisymmetric components of the mobility and diffusivity tensors, respectively, and
\begin{equation}
    \boldsymbol{\epsilon} =
    \begin{pmatrix}
        0 & 1 \\ -1 & 0
    \end{pmatrix}
\end{equation}
is the Levi-Civita tensor in two dimensions.
We allow the mobility tensor to have an antisymmetric component as well, which is often assumed in odd-diffusive systems.
For example, in an equilibrium environment, the fluctuation--dissipation relation, $\mathbf{D} =k_\mathrm{B} T \boldsymbol{\mu}$, with Boltzmann constant $k_\mathrm{B}$ and the environment temperature $T$, implies that a nonzero $\kappa_D$ necessarily accompanies a nonzero $\kappa_\mu$.
In the present work, we treat $\boldsymbol{\mu}$ and $\mathbf{D}$ independently in order to clearly distinguish their respective roles.

It is evident from Eq.~\eqref{eq:general_FP_eq} that the probability density $P(\boldsymbol{x},t)$ is independent of $\kappa_D$, since $\boldsymbol{\nabla}_n^\mathrm{T} \boldsymbol{\epsilon} \boldsymbol{\nabla}_n=0$. Thus, the antisymmetric part of the diffusion tensor contributes to the probability current but not to the time evolution of the density itself. The nontrivial question is whether this independence also extends to multipoint statistics.

To address this question, we now consider arbitrary state-dependent observables
$A_\alpha(\boldsymbol{x}(t_\alpha))$
evaluated at arbitrary times $t_\alpha$.
We derive that if the initial distribution $P_0(\boldsymbol{x}) \equiv P(\boldsymbol{x},0)$ is independent of $\kappa_D$, then
\begin{equation}\label{eq:main}
\partial_{\kappa_D}
\Big\langle
\Phi\left(
A_1(\boldsymbol{x}(t_1)),\ldots,
A_K(\boldsymbol{x}(t_K))
\right)
\Big\rangle
=0,
\end{equation}
where $\Phi(A_1,\ldots,A_K)$ is an arbitrary function of the observables.
This is the central result of the paper.

Choosing $\Phi = \prod_\alpha A_\alpha$ gives arbitrary multipoint correlation functions, defined as
\begin{equation}
    C_{\boldsymbol{A}} (\boldsymbol{t}) = \left \langle
    \prod_{\alpha=1}^K A_\alpha\left (\boldsymbol{x}(t_\alpha) \right)
    \right \rangle
\end{equation}
with $\boldsymbol{A}=(A_1,A_2,\ldots,A_K)^\mathrm{T}$ and $\boldsymbol{t}=(t_1,t_2,\ldots,t_K)^\mathrm{T}$.
Choosing $\Phi = \prod_\alpha \delta(a_\alpha - A_\alpha)$ gives the joint probability density of the measured observables.
This result implies that the statistics of arbitrary state-dependent observables remain completely unaffected by the odd diffusivity $\kappa_D$.

\section{Derivation}
\label{sec:derivation}

The key idea of the derivation is to embed the observables into an extended state space by introducing auxiliary variables.
For each observable $A_\alpha$, we introduce an auxiliary variable
$q_\alpha(t)$ that measures the change in $A_\alpha$ relative to its value at
$t_\alpha$ for $t > t_\alpha$ and vanishes otherwise, given by
\begin{equation}
    q_\alpha(t)
    =
    \begin{cases}
        0,
        & t<t_\alpha,\\[1ex]
        A_\alpha(\boldsymbol{x}(t))
        -
        A_\alpha(\boldsymbol{x}(t_\alpha)),
        & t\ge t_\alpha.
    \end{cases}
\end{equation}
Its time derivative is written as
\begin{equation}
    \dot q_\alpha(t)
    =
    \Theta(t-t_\alpha)
    \sum_n
    \boldsymbol{\nabla}_n^\textrm{T} A_\alpha(\boldsymbol{x}(t))
    \circ
    \dot{\boldsymbol x}_n(t),
\end{equation}
where $\circ$ represents the Stratonovich product and $\Theta(t)$ is the Heaviside step function.

The time evolution of the joint probability density $P(\boldsymbol{x},\boldsymbol{q},t)$ of the particle positions and the auxiliary variables is then given by~\cite{park2021thermodynamic}
\begin{equation}\label{eq:ex_FP}
    \dot{P} (\boldsymbol{x},\boldsymbol{q},t)
    = -\sum_n \widetilde{\boldsymbol{\nabla}}^\mathrm{T}_{n}
    \left( \boldsymbol{\mu} \boldsymbol{f}_n(\boldsymbol{x},t) - \mathbf{D} \widetilde{\boldsymbol{\nabla}}_n \right)
    P(\boldsymbol{x},\boldsymbol{q},t),
\end{equation}
where
\begin{equation}\label{eq:tilted_nabla}
    \widetilde{\boldsymbol{\nabla}}_n
    = \boldsymbol{\nabla}_n
    + \sum_\alpha \Theta(t-t_\alpha)
    [\boldsymbol{\nabla}_n A_\alpha(\boldsymbol{x})] \partial_{q_\alpha}
\end{equation}
is the tilted gradient operator.
The square brackets $[\square]$ indicate that the gradient operator
$\boldsymbol{\nabla}_n$ inside the bracket acts only on the enclosed quantity.

All dependence on $\kappa_D$ in Eq.~\eqref{eq:ex_FP} appears only through $\kappa_D \widetilde{\boldsymbol{\nabla}}_n^{\mathrm T}\boldsymbol{\epsilon}\widetilde{\boldsymbol{\nabla}}_n$.
Using Eq.~\eqref{eq:tilted_nabla} and the product rule,
$\boldsymbol{\nabla}^\textrm{T}_n \boldsymbol{\epsilon} [\boldsymbol{\nabla}_n A_\alpha (\boldsymbol{x})] = [\boldsymbol{\nabla}_n^\textrm{T} \boldsymbol{\epsilon} \boldsymbol{\nabla}_n A_\alpha (\boldsymbol{x})] + [\boldsymbol{\nabla}_n A_\alpha (\boldsymbol{x})]^\textrm{T} \boldsymbol{\epsilon}^\textrm{T} \boldsymbol{\nabla}_n$, we expand the operator
\begin{equation}
    \widetilde{\boldsymbol{\nabla}}_n^{\mathrm T}\boldsymbol{\epsilon}\widetilde{\boldsymbol{\nabla}}_n
    = \boldsymbol{\nabla}_n^{\mathrm T}\boldsymbol{\epsilon}\boldsymbol{\nabla}_n +
    \sum_\alpha
    \Theta(t-t_\alpha)
    [\boldsymbol{\nabla}^\textrm{T}_n \boldsymbol{\epsilon} \boldsymbol{\nabla}_n A_\alpha(\boldsymbol{x})]
    \partial_{q_\alpha} +
    \sum_{\alpha,\beta} \boldsymbol{\mathcal{A}}_{\alpha}^{\mathrm T} (\boldsymbol{x})
    \boldsymbol{\epsilon}\boldsymbol{\mathcal{A}}_{\beta} (\boldsymbol{x})
    \partial_{q_\alpha}\partial_{q_\beta},
\end{equation}
where $\boldsymbol{\mathcal{A}}_{\alpha} (\boldsymbol{x})=\Theta(t-t_\alpha)[\boldsymbol{\nabla}_n A_\alpha (\boldsymbol{x})]$.
Since $\boldsymbol{\nabla}_n^{\mathrm T}\boldsymbol{\epsilon}\boldsymbol{\nabla}_n=0$ and the summand in the last term is antisymmetric under the interchange of $\alpha$ and $\beta$, we arrive at
\begin{equation}
    \widetilde{\boldsymbol{\nabla}}_n^{\mathrm T}
    \boldsymbol{\epsilon}
    \widetilde{\boldsymbol{\nabla}}_n=0.
\end{equation}
The evolution equation for $P(\boldsymbol{x},\boldsymbol{q},t)$ is therefore independent of $\kappa_D$. Since the initial condition $P_0(\boldsymbol{x})\delta(\boldsymbol{q})$ is also independent of $\kappa_D$, the joint probability density $P(\boldsymbol{x},\boldsymbol{q},t)$ is independent of $\kappa_D$ at all later times.

Using $A_\alpha(\boldsymbol{x} (t_\alpha)) = A_\alpha(\boldsymbol{x}(t)) - q_\alpha(t)$, one obtains
\begin{equation}
    \left\langle \Phi\left(A_1 (\boldsymbol{x}(t_1)),\ldots,A_K(\boldsymbol{x}(t_K)) \right ) \right\rangle
    = \int d\boldsymbol{x} d\boldsymbol{q}
    \Phi(A_1(\boldsymbol{x}) - q_1,\ldots,A_K(\boldsymbol{x}) - q_K) P(\boldsymbol{x}, \boldsymbol{q}, t)
\end{equation}
for any $t > \max_\alpha t_\alpha$,
which immediately implies the main result in Eq.~\eqref{eq:main}.

\section{Physical meaning and consequences}
\label{sec:consequence}

This result provides a clear physical interpretation of the anomalous behaviors of multipoint state-dependent observables in odd-diffusive systems.
The antisymmetric diffusion tensor contributes to the probability flux, generating a Lorentz flux
\begin{equation}
    \boldsymbol{J}_n^\textrm{odd} (\boldsymbol{x},t)
    = - \kappa_D \boldsymbol{\epsilon} \boldsymbol{\nabla}_n P(\boldsymbol{x},t).
\end{equation}
The result shows that this Lorentz flux does not contribute to the statistics of state-dependent observables evaluated at arbitrary times.
Consequently, previously reported anomalous behavior must originate solely from the antisymmetric part of the mobility tensor.

A further implication is that the expectation values of
arbitrary functions of state-dependent observables can be
evaluated by setting $\kappa_D=0$.
This observation substantially simplifies both analytical and numerical studies of odd-diffusive systems, because a direct treatment of odd-diffusive dynamics is notoriously difficult.

The Fokker--Planck equation with an antisymmetric diffusion tensor corresponds to a stochastic process driven by non-white Gaussian noise $\boldsymbol{\zeta}_n (t)$ whose correlation structure is given by~\cite{chun2018emergence}
\begin{equation}
    \langle \boldsymbol{\zeta}_n(t) \boldsymbol{\zeta}_m^\mathrm{T} (t') \rangle
    = \left \{
    \begin{matrix}
        2 \mathbf{D} \delta_{n,m} \delta(t-t'), &
        \textrm{for } t>t', \\
        2 \mathbf{D}^\textrm{T} \delta_{n,m} \delta(t-t'), &
        \textrm{for } t<t'.
    \end{matrix}
    \right .
\end{equation}
The standard analytical and numerical techniques, formulated for white-noise Langevin dynamics, cannot be directly applied to such processes.
For this reason, one instead introduces an auxiliary dynamics that retains a conventional Langevin form and reduces to the original odd-diffusive dynamics in an appropriate limit.

For example, for systems in equilibrium environments satisfying the fluctuation--dissipation relation, the auxiliary dynamics is governed by~\cite{chun2018emergence,park2021thermodynamic}
\begin{equation}\label{eq:axu_Langevin}
    m \boldsymbol{\mu} \dot{\tilde{\boldsymbol{v}}}_n (t)= \boldsymbol{\mu} \boldsymbol{f}_n (\tilde{\boldsymbol{x}}(t),t) - \tilde{\boldsymbol{v}}_n (t)+ \sqrt{2 D_0} \boldsymbol{\xi}_n(t),
\end{equation}
with $\tilde{\boldsymbol{v}}_n (t) = \dot{\tilde{\boldsymbol{x}}}_n (t)$, a fictitious mass $m$, and white Gaussian noise $\boldsymbol{\xi}_n (t)$ characterized by $\langle \boldsymbol{\xi}_n(t) \boldsymbol{\xi}_{n'}^\mathrm{T} (t') \rangle = \mathbf{I}\delta_{n,n'}\delta(t-t')$.
The original odd-diffusive dynamics is recovered after taking the singular limit $m\to0$. Consequently, analytical and numerical calculations must first be performed in the auxiliary dynamics and subsequently extrapolated to the small-mass limit.
However, Eq.~\eqref{eq:main} bypasses this difficulty, yielding the same result directly from the dynamics with $\kappa_D=0$ and eliminating the need for an extrapolation in $m$.

Beyond this practical advantage, the result also has a more fundamental implication. Odd diffusivity does not alter any universal relation involving only state-dependent observables evaluated at multiple times. Therefore, every such relation that holds for ordinary Langevin dynamics driven by white Gaussian noise remains valid in odd-diffusive systems without modification.
Conversely, it implies that the presence of odd diffusivity cannot be inferred from any indicator constructed solely from state-dependent observables evaluated at multiple times. Detecting odd diffusivity therefore requires observables that explicitly probe probability currents or other quantities beyond state-dependent statistics.

\section{Examples}
\label{sec:examples}

We validate our results through several examples. First, we test the predicted independence from $\kappa_D$ for the mean back relaxation (MBR)~\cite{muenker2024accessing} of a Brownian particle confined in a quartic potential. We then show how this property can be exploited to obtain more accurate numerical estimates for the hard-core colliding particle system studied in Ref.~\cite{kalz2022collisions}. Finally, using a Brownian particle subject to linear forces, we verify that both the nonlinear fluctuation--dissipation relation (NL-FDR)~\cite{engbring2023nonlinear} and the generalized Green--Kubo relation~\cite{pavliotis2015stochastic} remain valid in odd-diffusive systems.
Further details are provided in the Appendices.

Throughout this section, we focus on equilibrium environments where the mobility tensor and diffusion tensor are expressed as
\begin{equation}\label{eq:FDR}
    \boldsymbol{\mu} = \frac{1}{k_\textrm{B}T} \mathbf{D},
    \quad
    \mathbf{D} = D_0 \mathbf{I} + \kappa \boldsymbol{\epsilon}.
\end{equation}
For convenience, we also consider redefined parameters $\gamma$ and $B$ given by
\begin{equation}\label{def:gamma_B}
    \gamma \equiv \frac{k_\textrm{B} T D_0}{D_0^2 + \kappa^2},
    \qquad
    B \equiv \frac{k_\textrm{B} T \kappa}{D_0^2 + \kappa^2}.
\end{equation}
For simplicity, we set $k_\textrm{B}=1$ and also use $\gamma=T=1$ for all examples.
The units of length and time are introduced separately in each example. 

\subsection{Mean back relaxation}

\begin{figure}
        \centering
        \includegraphics[width=0.5\textwidth]
        {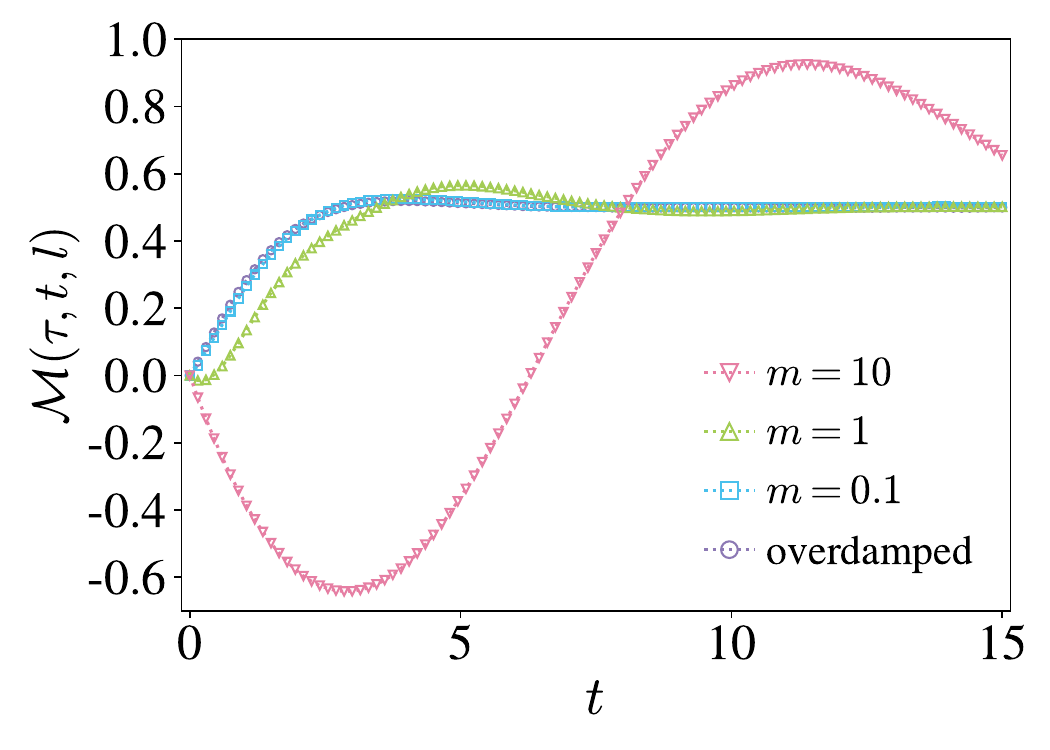}
		\caption{
        Mean back relaxation $\mathcal{M}(\tau,t,l)$ for a Brownian particle in a quartic potential. 
        Symbols show the results obtained from the auxiliary underdamped dynamics for $m = 10$ (magenta downward triangles), $1$ (green upward triangles), and $0.1$ (blue squares), and from the overdamped dynamics with $\kappa_D=0$ (violet circles).
        }
        \label{fig:MBR}
\end{figure}

The MBR is a three-point measure that serves as an indicator of time-reversal symmetry breaking~\cite{muenker2024accessing,knotz2024mean}. For a scalar coordinate $x$, it is defined as
\begin{equation}
    \mathcal{M} (\tau, t, l) = \left \langle
    - \frac{x(t) - x(0)}{x(0) - x(-\tau)}
    \vartheta_l(|x(0) - x(-\tau)|)
    \right \rangle,
\end{equation}
where
\begin{equation}
    \vartheta_l(r) = \frac{\theta(r - l)}
    {\langle \theta(r - l) \rangle}
\end{equation}
is a normalized masking factor that excludes trajectories with small denominators $|x(0) - x(-\tau)|$.

We consider a single particle confined by a quartic potential
\begin{equation}
    U(\boldsymbol{x}) = \frac{k'}{4}(x^4 + y^4).
\end{equation}
The length and time are expressed in units of
\begin{equation}
    r^\textrm{I} \equiv \left( \frac{T}{k'} \right)^{\frac{1}{4}}
    \quad \textrm{and}\quad
    \tau^\textrm{I} = \frac{\gamma}{\sqrt{k' T}}.
\end{equation}
The selected parameters are $B = k'= 1$, $\tau=2$, and $l=0.2$, while the mass is varied over $m=0.1$, $1$, and $10$.

We compare the MBR for the auxiliary underdamped dynamics with that for the ordinary overdamped Langevin dynamics corresponding to $\kappa_D=0$.
The time-dependent behavior of the MBR obtained by numerical integration is shown in Fig.~\ref{fig:MBR}.
As the fictitious mass $m$ decreases, the MBR obtained from the auxiliary dynamics approaches that obtained from the ordinary Langevin dynamics, as predicted by the main result.

\subsection{Diffusion coefficient enhanced by collisions}

\begin{figure}
    \centering
    \includegraphics[width=0.5\textwidth]
    {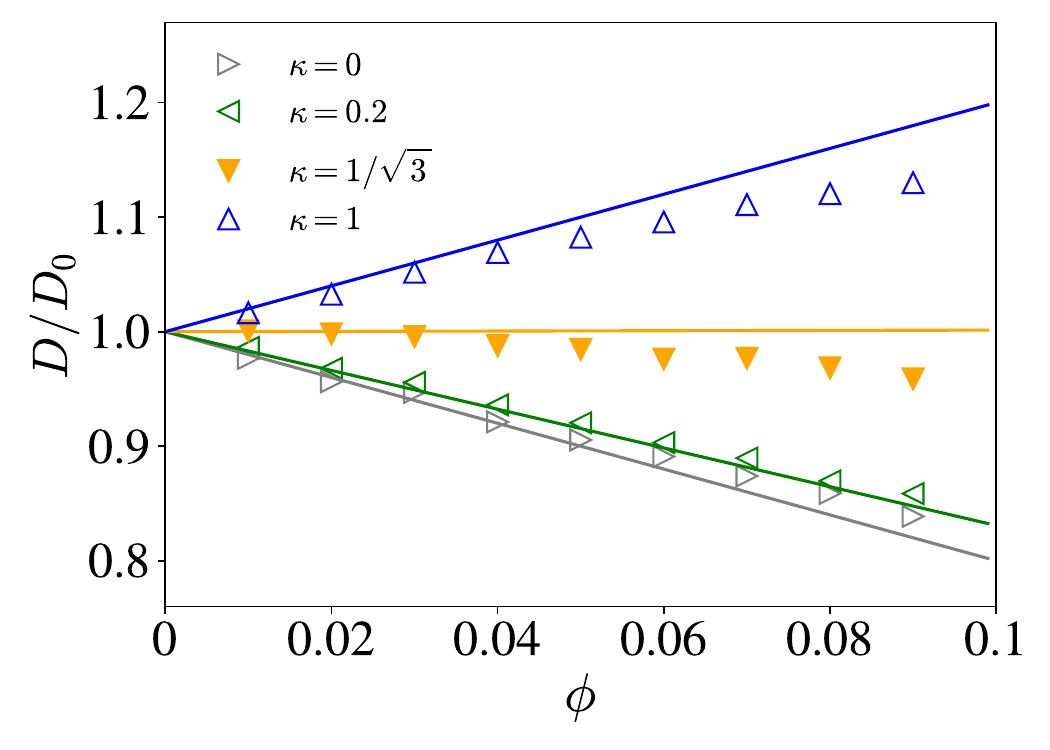}
		\caption{
        Effective diffusion coefficient $D$ for hard-core repulsive particles obtained from the overdamped dynamics ($\kappa_D=0$). Symbols show numerical results as a function of the packing fraction $\phi$ for $\kappa=\kappa_\mu=0$ (gray right-pointing triangles), $0.2$ (green left-pointing triangles), $1/\sqrt{3}$ (orange downward triangles), and $1$ (blue upward triangles). Colored solid lines show the corresponding analytical predictions in the small-$\phi$ limit.
        }
		\label{fig:hardcore}
	\end{figure}

We next revisit a hard-core repulsive particle system previously studied in Ref.~\cite{kalz2022collisions}. The system consists of $N=200$ identical particles of diameter $\sigma$ interacting through hard-core repulsion, confined in a square domain of side length $L$ with periodic boundary conditions. Thus the packing fraction is given by
\begin{equation}
    \phi = \frac{N \pi \sigma^2}{4 L^2}.
\end{equation}
The effective diffusion coefficient of a tracer particle is defined as
\begin{equation}
    D = \lim_{t\rightarrow\infty} \frac{\langle |\boldsymbol{x}_n(t) - \boldsymbol{x}_n(0)|^2 \rangle}{4t}.
\end{equation}
To enable a direct comparison with Ref.~\cite{kalz2022collisions}, we use the same parameter values.

Figure~\ref{fig:hardcore} shows the averaged $D$ as a function of the packing fraction $\phi$ for several values of the odd coefficient $\kappa$.
The symbols represent numerical results obtained by integrating the ordinary Langevin equation. The solid curves denote analytical predictions derived in the low-density limit~\cite{kalz2022collisions}. For small packing fractions, excellent agreement is observed between the numerical and analytical results.

This agreement was not clearly visible in the original study, where $D$ was computed using the auxiliary underdamped dynamics. In particular, the numerical results reported there appeared to show a diffusion enhancement at $\kappa=\kappa_{\mathrm c}\equiv 1/\sqrt{3}$, which is inconsistent with the theoretical prediction that the effective diffusion coefficient can exceed $D_0$ only for $\kappa>\kappa_{\mathrm c}$. By contrast, the present numerical results are fully consistent with the analytical prediction because they are free from the finite-$m$ errors inherent in the auxiliary-dynamics approach.

\subsection{Nonlinear fluctuation--dissipation relation}

The NL-FDR was originally proposed as a necessary condition for Markovian dynamics, and thus has been used to identify non-Markovian processes~\cite{engbring2023nonlinear}.
The key quantity of interest is an indicator defined by
\begin{equation}
    z(\boldsymbol{x}) \equiv
    \frac{P^\epsilon (\boldsymbol{x})}{P^\textrm{s} (\boldsymbol{x})} - 1
\end{equation}
where $P^\textrm{s}(\boldsymbol{x})$ is the steady-state distribution, whereas $P^\epsilon(\boldsymbol{x})$ is an arbitrary distribution regarded as the steady-state distribution in the presence of a perturbation.
The NL-FDR shows a relation between the relaxation of the average value of $z(\boldsymbol{x})$ and its autocorrelation in the steady state, which reads
\begin{equation}
    \langle z(\boldsymbol{x}(t)) \rangle^\epsilon
    = C_{z,z}^\textrm{s} (t,0),
\end{equation}
where 
$\langle \cdots \rangle^\epsilon$ denotes an average over a relaxation process evolving from $P^\epsilon(\boldsymbol{x})$, and
$C_{z,z}^\textrm{s} (t,0)$ is the autocorrelation function of $z(\boldsymbol{x})$ in the steady state.
The superscript $\textrm{s}$ indicates that the corresponding quantity is obtained in the stationary state.
Our results imply that this relation holds for odd-diffusive systems.
This relation has previously been derived only for Langevin dynamics driven by white Gaussian noise.
Our result therefore identifies odd-diffusive systems as a nontrivial class of stochastic processes for which the NL-FDR remains valid.

\begin{figure}
    \centering
    \includegraphics[width=\textwidth]
    {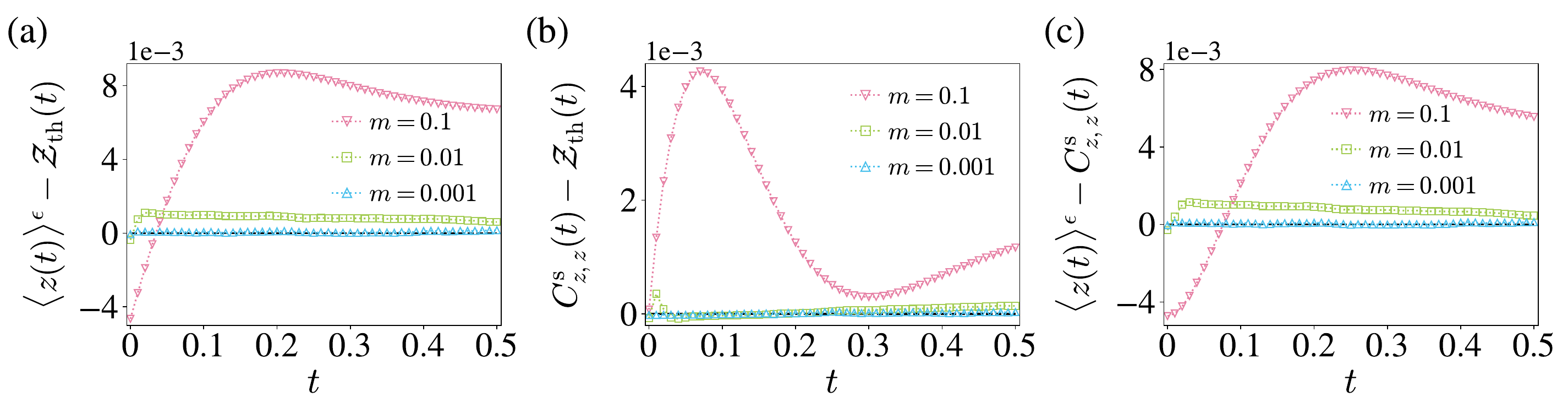}
		\caption{
        Verification of the nonlinear fluctuation--dissipation relation in an odd-diffusive system. Panels (a) and (b) show the deviations of $\langle z(\boldsymbol{x}(t))\rangle^\epsilon$ and $C_{z,z}^{\mathrm s}(t,0)$, respectively, from the analytical prediction $\mathcal{Z}_{\rm th}(t)$ obtained from the ordinary overdamped Langevin dynamics. Panel (c) shows the difference $\langle z(\boldsymbol{x}(t))\rangle^\epsilon - C_{z,z}^{\mathrm s}(t,0)$. Symbols show numerical results obtained from the auxiliary dynamics for $m=0.1$ (magenta downward triangles), $m=0.01$ (green squares), and $m=0.001$ (blue upward triangles).
        }
		\label{fig:NLFDR}
	\end{figure}

We examine this relation in a Brownian particle confined in a harmonic potential with a linear perturbation, where the force is given by
\begin{equation}\label{def:lin_force}
    \boldsymbol{f}(\boldsymbol{x})
    = -
    \begin{pmatrix}
    k & -\epsilon \\
    \epsilon & k
    \end{pmatrix}
    \boldsymbol{x}.
\end{equation}
The length and time units are given by
\begin{equation}\label{eq:units_II}
    r^\textrm{II} \equiv \sqrt{\frac{T}{k}}
    \quad \textrm{and} \quad
    \tau^\textrm{II} = \frac{\gamma}{k}
\end{equation}
The selected parameters are $B=k=1$ and $\epsilon=0.5$, while the mass is varied over $m=0.001$, $0.01$, and $0.1$.

For the ordinary Langevin case, the NL-FDR holds trivially, and we can obtain the analytical expression
\begin{equation}
    \langle z (\boldsymbol{x}(t)) \rangle^\epsilon
    =
    \frac{\epsilon^2 B^2}{e^{t/\tau} (k \gamma + \epsilon B)^2 -\epsilon^2 B^2}
    \equiv \mathcal{Z}_\textrm{th} (t)
\end{equation}
with $\tau = T/(2kD_0)$.
Thus it also satisfies
$C_{z,z}^\textrm{s}(t,0) = \mathcal{Z}_\textrm{th}(t)$.
We compare $\langle z(\boldsymbol{x}(t))\rangle^\epsilon$ and $C_{z,z}^\mathrm{s}(t,0)$ numerically obtained in the auxiliary dynamics and the analytical result obtained in the ordinary Langevin dynamics.
Figures~\ref{fig:NLFDR}(a) and \ref{fig:NLFDR}(b)
show the deviations of
$\langle z(\boldsymbol{x}(t))\rangle^\epsilon$
and $C_{z,z}^\mathrm{s}(t,0)$, respectively, from
$\mathcal{Z}_\textrm{th}(t)$ for
$m=0.1$, $0.01$, and $0.001$.
They show that both quantities converge to the analytical result obtained in the ordinary Langevin dynamics as $m$ decreases.
We also evaluate the deviation from the NL-FDR in the auxiliary underdamped dynamics, as shown in Fig.~\ref{fig:NLFDR}(c).
The deviation gradually vanishes as $m$ decreases, consistent with the prediction that the NL-FDR remains valid in odd-diffusive systems.
These results are consistent with our main result and verify that the NL-FDR remains valid in odd-diffusive systems.

\subsection{Generalized Green--Kubo relation}

   \begin{figure*}
    \centering
    \includegraphics[width=\textwidth]
    {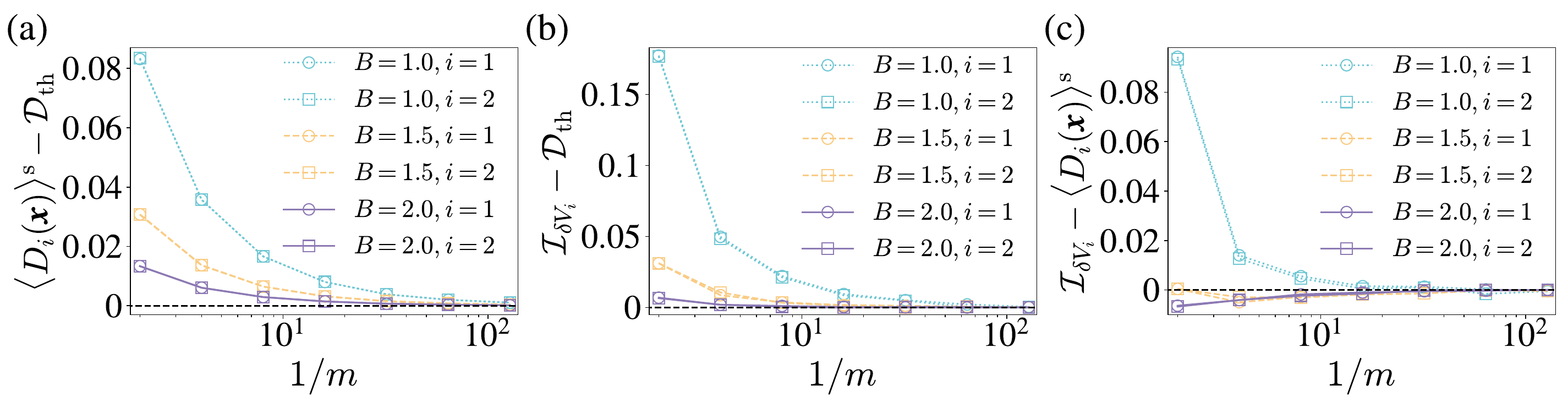}
		\caption{
        Verification of the generalized Green--Kubo relation. Panels (a) and (b) show the deviations of the steady-state average of the generalized diffusion coefficient $\langle D_i(\boldsymbol{x})\rangle^\mathrm{s}$ and the integrated generalized drift correlation $\mathcal{I}_{\delta V_i}$, respectively, from the analytical prediction $\mathcal{D}_\textrm{th}$ obtained from the overdamped Langevin dynamics. Panel (c) shows the difference $\mathcal{I}_{\delta V_i}-\langle D_i(\boldsymbol{x})\rangle^\mathrm{s}$. Symbols show numerical results obtained from the auxiliary dynamics. Cyan symbols connected by dotted lines, yellow symbols connected by dashed lines, and violet symbols connected by solid lines correspond to $B=1$, $1.5$, and $2$, respectively. Circles and squares correspond to the first and second coordinates ($i=1$ and $2$), respectively.
        }
		\label{fig:GGKR}
	\end{figure*}

As a final example, we consider the generalized Green--Kubo relation~\cite{pavliotis2015stochastic}.
It extends the Green--Kubo relation to generalized
drift and diffusion coefficients for arbitrary functions
$f(\boldsymbol{x})$ and $g(\boldsymbol{x})$, which are
defined by
\begin{equation}
    V^f(\boldsymbol{x}) = \lim_{dt \rightarrow0} \frac{\langle df(\boldsymbol{x}) \rangle}{dt},
    \qquad
    D^{f,g}(\boldsymbol{x}) = \lim_{dt\rightarrow0} \frac{\langle df(\boldsymbol{x}) dg(\boldsymbol{x}) \rangle}{dt},
\end{equation}
where $df(\boldsymbol{x}) = f(\boldsymbol{x}(t+dt))-f(\boldsymbol{x}(t)) |_{\boldsymbol{x}(t)=\boldsymbol{x}}$.
Then the generalized Green--Kubo relation reads
\begin{equation}
    \langle D^{f,g}(\boldsymbol{x}) \rangle^\textrm{s}
    = \int_0^\infty dt' \left ( C_{\delta V^{f},\delta V^{g}}^\textrm{s}(t',0) + C_{\delta V^{g},\delta V^{f}}^\textrm{s}(t',0)  \right ),
\end{equation}
where
$C^\textrm{s}_{\delta V^f,\delta V^g}(t,0)$
is the centered time correlation between
$V^f(\boldsymbol{x}(t))$ and
$V^g(\boldsymbol{x}(0))$, and
$\delta V^f(\boldsymbol{x})
\equiv V^f(\boldsymbol{x})
-\langle V^f(\boldsymbol{x})\rangle^\textrm{s}$
denotes the deviation from the steady-state average.
The validity of this generalized Green--Kubo relation has not been proven for odd-diffusive systems, although specific forms of the relation have been derived for odd-diffusive systems~\cite{yasuda2022time}.
Our results imply that for arbitrary functions $f$ and $g$, this generalized Green--Kubo relation holds for odd-diffusive systems.

For numerical verification, we employ $f(\boldsymbol{x})=g(\boldsymbol{x})=x_i^2/2$  for $i=1, 2$, where $x_1 = x$ and $x_2=y$.
The generalized drift and diffusion coefficients are then given by
\begin{equation}
    V_i (\boldsymbol{x}) \equiv V^{x_i^2/2}(\boldsymbol{x}) = [\boldsymbol{\mu} \boldsymbol{f}(\boldsymbol{x})]_i x_i + D_0
\end{equation}
and
\begin{equation}
    D_i(\boldsymbol{x})\equiv D^{x_i^2/2,x_i^2/2}(\boldsymbol{x})
    = 2 x_i^2 D_0,
\end{equation}
respectively.
The generalized Green--Kubo relation is then simplified as
\begin{equation}
    \langle D_i(\boldsymbol{x}) \rangle^\textrm{s}
    = \mathcal{I}_{\delta V_i},
\end{equation}
where
\begin{equation}
     \mathcal{I}_{\delta V_i}\equiv 2 \int_0^\infty dt' C_{\delta V_i, \delta V_i}^\textrm{s}(t',0).
\end{equation}

We consider the system with the same linear external force given in Eq.~\eqref{def:lin_force}.
The length and time units are the same as those in
Eq.~\eqref{eq:units_II}.
The selected parameters are $k=1$ and $\epsilon=0.5$,
while $B$ is set to $1$, $1.5$, and $2$, and $m$ is varied.

In the ordinary Langevin dynamics, we obtain the analytical form
\begin{equation}
    \langle D_i (\boldsymbol{x}) \rangle^\textrm{s}
    = 
    \frac{2 (\gamma  T)^2}{(\gamma^2+B^2)(k \gamma  +\epsilon B)} \equiv \mathcal{D}_\textrm{th}.
\end{equation}

We compare this analytical solution with each quantity obtained in the auxiliary dynamics.
The deviations of
$\langle D_i(\boldsymbol{x})\rangle^\textrm{s}$
and $\mathcal{I}_{\delta V_i}$
are shown in Figs.~\ref{fig:GGKR}(a) and
\ref{fig:GGKR}(b), respectively, while their difference
is shown in Fig.~\ref{fig:GGKR}(c).
The results consistently demonstrate that all deviations gradually vanish as $m$ decreases.
These results verify the generalized Green--Kubo relation in odd-diffusive systems, as expected from our main result.

\section{Conclusion}
\label{sec:conclusion}

We have shown that the Lorentz flux generated by the antisymmetric part of the diffusion tensor does not affect any multipoint statistics constructed from state-dependent observables. The result is an extension of the observation that the one-time density evolution is independent of odd diffusivity: it applies to arbitrary functions of arbitrary observables evaluated at arbitrary times, and therefore includes correlation functions and full joint probability distributions of measured state variables.

The theorem clarifies the origin of anomalous correlations in odd-diffusive systems. Effects such as collision-enhanced self-diffusion and oscillatory autocorrelation functions do not originate from the antisymmetric diffusion coefficient itself but instead arise from transverse forces and nonreciprocal effective interactions produced by the antisymmetric mobility.

This distinction has practical consequences. Whenever the quantity of interest is a state-observable statistic, the antisymmetric diffusion coefficient can be set to zero and the calculation can be performed using an ordinary white-noise Langevin equation while retaining odd mobility. This avoids the auxiliary underdamped dynamics and the associated small-mass extrapolation normally required to represent non-white noise. Revisiting the hard-core repulsive particle example illustrates that this direct route can remove finite-mass artifacts and recover the low-density theoretical prediction more clearly.

Because odd diffusivity does not modify the multipoint statistics of state-dependent observables, relations such as the nonlinear fluctuation--dissipation relation and the generalized Green--Kubo relation remain valid without modification in odd-diffusive systems. Conversely, the theorem identifies what is needed to observe odd diffusivity: it is invisible to any measurement constructed solely from state-dependent observables.
One must measure currents or other quantities that explicitly probe probability flow rather than only the history of state variables. This provides a clear criterion for distinguishing the contributions of odd mobility and odd diffusivity.

\appendix

\section{Numerical simulation}

In all numerical examples, we consider the equilibrium case satisfying the fluctuation--dissipation relation in Eq.~\eqref{eq:FDR}.  With the parameters defined in Eq.~\eqref{def:gamma_B}, the auxiliary underdamped dynamics is written as
\begin{equation}\label{eq:auxiliary_dynamics_app}
    m\dot{\boldsymbol{v}}(t)
    = \boldsymbol{f}(\boldsymbol{x}(t))
    - \mathbf{G}\boldsymbol{v}(t)
    + \sqrt{2\gamma T}\,\boldsymbol{\xi}(t),
\end{equation}
where
\begin{equation}
    \mathbf{G}
    \equiv
    \begin{pmatrix}
        \gamma & -B \\
        B & \gamma
    \end{pmatrix}.
\end{equation}
This equation is numerically integrated using the Euler--Maruyama method.
The corresponding ordinary overdamped dynamics with $\kappa_D=0$ and $\kappa_\mu=\kappa$ is
\begin{equation}\label{eq:ordinary_overdamped_app}
    \dot{\boldsymbol{x}}(t)
    = \mathbf{G}^{-1}\boldsymbol{f}(\boldsymbol{x}(t))
    + \sqrt{\frac{2\gamma T}{\gamma^2+B^2}}\,\boldsymbol{\xi}(t).
\end{equation}
This overdamped equation is directly simulated in the first two examples.  For the examples with linear forces, we use the analytical solution of the ordinary overdamped dynamics.

\subsection{Example I: MBR}

In the first example, we consider a single-particle system, $N=1$, where the force is given by
\begin{equation}
    \boldsymbol{f}(x,y)=-\boldsymbol{\nabla} U(x,y)
\end{equation}
with
\begin{equation}
    U(x,y)=\frac{k'}{4}(x^4+y^4).
\end{equation}
The auxiliary underdamped dynamics is initialized from the equilibrium Boltzmann distribution
\begin{equation}
    P^\textrm{eq}(\boldsymbol{x},\boldsymbol{v})
    =P^\textrm{eq}(\boldsymbol{x})P^\textrm{eq}(\boldsymbol{v}),
\end{equation}
where
\begin{equation}
    P^\textrm{eq}(\boldsymbol{x})
    =\frac{1}{\mathcal N}
    \exp\left(-\frac{U(x,y)}{T}\right)
\end{equation}
and
\begin{equation}
    P^\textrm{eq}(\boldsymbol{v})
    =\frac{m}{2\pi T}
    \exp\left(-\frac{m\boldsymbol{v}^\textrm{T}\boldsymbol{v}}
    {2T}\right).
\end{equation}
Here $\mathcal N$ is the normalization constant of the position distribution.  For the ordinary overdamped dynamics, only the initial position is sampled from $P^\textrm{eq}(\boldsymbol{x})$.

The equilibrium position is generated by using a Gamma-distributed random variable.  We first sample $w$ from
\begin{equation}
    p(w)=\frac{1}{\theta^\alpha \Gamma(\alpha)}w^{\alpha-1}e^{-\frac{w}{\theta}},
    \qquad
    \alpha=\frac14,
    \qquad
    \theta=\frac{4T}{k'},
\end{equation}
and then set
\begin{equation}
    z=\pm w^{1/4},
\end{equation}
where the two signs are chosen with equal probability.  This procedure gives
\begin{equation}
    P(z)\propto
    \exp\left[-\frac{k'z^4}{4T}\right].
\end{equation}
The integration time step is fixed at $\Delta t=10^{-3}$.  For each parameter set, $10^7$ independent trajectories are generated.  The first 150 integration steps are discarded before evaluating the MBR $\mathcal M(\tau,t,l)$.

\subsection{Example II: Diffusion coefficient}

In the second example, we directly simulate the ordinary overdamped Langevin dynamics for the colliding-particle model discussed in Ref.~\cite{kalz2022collisions}.  The repulsive force acting on particle $i$ from particle $j$ is given by
\begin{equation}
    \boldsymbol{f}(\boldsymbol{x}_i,\boldsymbol{x}_j)
    =\varepsilon
    \left(\frac{\sigma}{r_{i,j}}\right)^\alpha
    \Theta(\sigma_\textrm{c}-r_{i,j})
    (\boldsymbol{x}_i-\boldsymbol{x}_j),
\end{equation}
where
\begin{equation}
    r_{i,j}
    =\sqrt{(\boldsymbol{x}_i-\boldsymbol{x}_j)^\textrm{T}
    (\boldsymbol{x}_i-\boldsymbol{x}_j)}.
\end{equation}
Periodic boundary conditions are imposed in both spatial directions.  The box size $L$ is determined by the particle number and the packing fraction.  The parameters are chosen as $N=200$, $\varepsilon=100$, $\alpha=17$, $\sigma=1$, and $\sigma_\textrm{c}=1.01$.

Initially, the particle positions are sampled uniformly in the simulation box.  The overdamped equation is integrated with time step $\Delta t=10^{-5}$.  To avoid unstable initial configurations, the system is first evolved for 100 integration steps.  If the maximum repulsive force amplitude $\max_{i}|\sum_{j\neq i} \boldsymbol{f} (\boldsymbol{x}_i, \boldsymbol{x}_j)|$ exceeds 100 after this preliminary evolution, the initialization is discarded and repeated.

The particle positions are recorded every $t_\textrm{sample}$ integration steps, yielding the trajectory
\begin{equation}
    \Gamma=\{\boldsymbol{x}(t_0),\boldsymbol{x}(t_1),\ldots,
    \boldsymbol{x}(t_{\mathcal T})\},
\end{equation}
where
\begin{equation}
    t_a=a t_\textrm{sample}\Delta t.
\end{equation}
To estimate the effective diffusion coefficient, we divide the trajectory into overlapping segments
\begin{equation}
    \Gamma_k=\{\boldsymbol{x}(t_{wk}),\boldsymbol{x}(t_{wk+1}),\ldots,
    \boldsymbol{x}(t_{wk+\mathcal T_\textrm{w}})\},
\end{equation}
for
\begin{equation}
    k=0,\ldots,\mathcal K_\textrm{w},
    \qquad
    w=\frac{\mathcal T-\mathcal T_\textrm{w}}{\mathcal K_\textrm{w}}.
\end{equation}
We use $t_\textrm{sample} = 100$, $\mathcal T=19990$, $\mathcal T_\textrm{w}=10000$, and $\mathcal K_\textrm{w}=999$.
For each segment, the squared displacement is averaged over all particles, spatial coordinates, and 400 independent realizations.  The effective diffusion coefficient is obtained by fitting the last 80\% of the mean-square displacement to
\begin{equation}
    \left\langle (x(t)-x(0))^2\right\rangle
    =2D t+b,
\end{equation}
where $b$ is a fitting parameter.  The plotted value is obtained by averaging over all segments.

\subsection{Example III: NL-FDR}

In the third example, we consider a single-particle system, $N=1$, with a harmonic force
\begin{equation}
    \boldsymbol{f}(\boldsymbol{x})=-k\boldsymbol{x}.
\end{equation}
For the relaxation average $\langle z(\boldsymbol{x}(t))\rangle^\epsilon$, the auxiliary dynamics is initialized from
\begin{equation}\label{eq:ss_perturb_app}
    P^\epsilon(\boldsymbol{x},\boldsymbol{v})
    =\frac{1}{\sqrt{\det(2\pi\mathbf C^\epsilon)}}
    \exp\left(-\frac12
    \boldsymbol{u}^\textrm{T}(\mathbf C^\epsilon)^{-1}\boldsymbol{u}\right),
\end{equation}
where
\begin{equation}
    \boldsymbol{u}
    =\begin{pmatrix}
        \boldsymbol{x}\\
        \boldsymbol{v}
    \end{pmatrix}
\end{equation}
and the covariance matrix is
\begin{equation}\label{eq:aux_cov_app}
    \mathbf C^\epsilon
    =\frac{\gamma T}
    {\gamma(k\gamma+\epsilon B)-m\epsilon^2}
    \begin{pmatrix}
        \gamma & 0 & 0 & -\epsilon \\
        0 & \gamma & \epsilon & 0 \\
        0 & \epsilon & \frac{k\gamma+\epsilon B}{m} & 0 \\
        -\epsilon & 0 & 0 & \frac{k\gamma+\epsilon B}{m}
    \end{pmatrix}.
\end{equation}
For the steady-state autocorrelation $C^\textrm{s}_{z,z}(t,0)$, the initial distribution is taken to be $P^{\epsilon=0}(\boldsymbol{x},\boldsymbol{v})$.
Since $\langle z\rangle^\mathrm{s}=0$, $C_{z,z}^\mathrm{s}(t,0)$ and $C_{\delta z,\delta z}^\mathrm{s}(t,0)$ are identical. We evaluate the latter because it is numerically more convenient.
The time step is $\Delta t=10^{-3}$, $10^{-4}$, and $10^{-5}$ for $m=0.1$, $0.01$, and $0.001$, respectively.
For each value of $m$, $10^7$ samples are generated.

\subsection{Example IV: Generalized GKR}

In the final example, we consider a single-particle system, $N=1$, with a linear force in Eq.~\eqref{def:lin_force}.
The initial distribution of the auxiliary dynamics is the steady-state Gaussian distribution in Eq.~\eqref{eq:ss_perturb_app}, with the covariance matrix in Eq.~\eqref{eq:aux_cov_app}.

The integrated correlation $\mathcal I_{\delta V_i}$ is evaluated from $10^7$ samples with time step $\Delta t=10^{-4}$.  In the simulation, the centered correlation $C^\textrm{s}_{\delta V_i,\delta V_j}(t,0)$ is accumulated at every integration step.  The integral is calculated over $3\times10^5$ integration steps, corresponding to the total time $t=30$.

The steady-state average $\langle D_i(\boldsymbol{x})\rangle^\textrm{s}$ is evaluated separately. Starting from the initial state,
$\langle D_i(\boldsymbol{x})\rangle^\textrm{s}$ is averaged over $10^7$ samples after every 1000 integration steps. This procedure is repeated ten times, and the resulting estimates are averaged.

\section{Analytical solutions}\label{sec:analytic_linear_app}

\subsection{Steady-state and propagator in the linear models}

The ordinary overdamped Langevin equation with a force linear in position is written as
\begin{equation}\label{eq:linear_ou_app}
    \dot{\boldsymbol{x}}(t)
    =-\mathbf A\boldsymbol{x}(t)+\sqrt{2D_0}\,\boldsymbol{\xi}(t).
\end{equation}
The steady-state distribution is a Gaussian distribution,
\begin{equation}\label{expr:ss_dist_app}
    P^\textrm{s}(\boldsymbol{x})
    =\frac{1}{\sqrt{\det(2\pi\mathbf C^\textrm{s})}}
    \exp\left(-\frac12
    \boldsymbol{x}^\textrm{T}\mathbf Q^\textrm{s}\boldsymbol{x}\right),
\end{equation}
where $\mathbf C^\textrm{s}$ is the steady-state covariance matrix and $\mathbf Q^\textrm{s}=(\mathbf C^\textrm{s})^{-1}$.  The covariance matrix is obtained from the Lyapunov equation
\begin{equation}
    \mathbf A\mathbf C^\textrm{s}
    +\mathbf C^\textrm{s}\mathbf A^\textrm{T}
    =2D_0\mathbf I.
\end{equation}

During the relaxation, the mean and covariance are given by
\begin{equation}
    \langle\boldsymbol{x}(t)\rangle
    =e^{-\mathbf A t} \langle\boldsymbol{x}(0)\rangle
    \equiv\mathbf E(t) \langle\boldsymbol{x}(0)\rangle
\end{equation}
and
\begin{equation}
    \mathbf C(t)
    =\mathbf C^\textrm{s}
    +\mathbf E(t)(\mathbf C(0)-\mathbf C^\textrm{s})
    \mathbf E^\textrm{T}(t),
\end{equation}
respectively.  The propagator is therefore
\begin{equation}\label{expr:propagator_app}
    P(\boldsymbol{x},t|\boldsymbol{x}_0,0)
    =\frac{1}{\sqrt{\det(2\pi\mathbf C^\textrm{p}(t))}}
    \exp\left(-\frac12
    (\boldsymbol{x}-\mathbf E(t)\boldsymbol{x}_0)^\textrm{T}
    \mathbf Q^\textrm{p}(t)
    (\boldsymbol{x}-\mathbf E(t)\boldsymbol{x}_0)\right),
\end{equation}
where
\begin{equation}
    \mathbf C^\textrm{p}(t)
    =\mathbf C^\textrm{s}-\mathbf E(t)\mathbf C^\textrm{s}\mathbf E^\textrm{T}(t),
    \qquad
    \mathbf Q^\textrm{p}(t)= ( \mathbf C^\textrm{p}(t) )^{-1}.
\end{equation}

For the linear force in Eq.~\eqref{def:lin_force}, the drift matrix of the ordinary overdamped dynamics is
\begin{equation}
    \mathbf A^\epsilon
    =\frac{1}{T}
    \begin{pmatrix}
        D_0 & \kappa \\
        -\kappa & D_0
    \end{pmatrix}
    \begin{pmatrix}
        k & -\epsilon \\
        \epsilon & k
    \end{pmatrix},
\end{equation}
and the diffusion matrix is $D_0\mathbf I$.  The steady-state covariance matrix is then
\begin{equation}\label{eq:linear_cov_app}
    \mathbf C^\epsilon
    =\frac{D_0 T}{kD_0+\epsilon\kappa}\mathbf I.
\end{equation}

\subsection{Analytical solution for NL-FDR}

For the NL-FDR example, the perturbed initial state is a Gaussian distribution with covariance matrix $\mathbf C^\epsilon$ in Eq.~\eqref{eq:linear_cov_app}.  The indicator variable is written as
\begin{equation}
    z(\boldsymbol{x})
    =A^z\exp\left(-\frac12
    \boldsymbol{x}^\textrm{T}\mathbf Q^z\boldsymbol{x}\right)-1,
\end{equation}
where
\begin{equation}
    A^z = \sqrt{\frac{\det(\mathbf{Q}^\epsilon)}{\det(\mathbf{Q}^0)}},
    \qquad
    \mathbf{Q}^z = \mathbf{Q}^\epsilon-\mathbf{Q}^0
\end{equation}
with
$\mathbf{Q}^\epsilon = (\mathbf{C}^\epsilon)^{-1}$
and
$\mathbf{Q}^0=\mathbf{Q}^{\epsilon=0}$.
It is useful to introduce the conditional average
\begin{equation}
    \langle z(\boldsymbol{x}(t))|\boldsymbol{x}_0\rangle
    \equiv
    \int d\boldsymbol{x}\,z(\boldsymbol{x})
    P(\boldsymbol{x},t|\boldsymbol{x}_0,0).
\end{equation}
By using the Gaussian product theorem, we obtain
\begin{equation}\label{eq:cond_ave_x0_app}
    \langle z(\boldsymbol{x}(t))|\boldsymbol{x}_0\rangle
    =A^\textrm{c}\exp\left(-\frac12
    \boldsymbol{x}_0^\textrm{T}\mathbf Q^\textrm{c}(t)\boldsymbol{x}_0\right)-1,
\end{equation}
with
\begin{equation}
    A^\textrm{c}
    =A^z\sqrt{\frac{\det (\mathbf Q^\textrm{p}(t))}
    {\det(\mathbf Q^\textrm{p}(t)+\mathbf Q^z)}}
\end{equation}
and
\begin{equation}
    \mathbf Q^\textrm{c}(t)
    =(\mathbf E^0(t))^\textrm{T}\mathbf Q^\textrm{p}(t)
    \left((\mathbf Q^\textrm{p}(t))^{-1}
    -(\mathbf Q^\textrm{p}(t)+\mathbf Q^z)^{-1}\right)
    \mathbf Q^\textrm{p}(t)\mathbf E^0(t),
\end{equation}
where $\mathbf E^0(t) = e^{-\mathbf A^0 t}$ and
\begin{equation}
    \mathbf{Q}^\textrm{p}(t)
    = \left (
    \mathbf{C}^0
     - \mathbf E^0(t) \mathbf{C}^0 (\mathbf E^0(t))^\textrm{T}
    \right)^{-1}
\end{equation}
are obtained in the unperturbed dynamics with
\begin{equation}
    \mathbf C^0
    = \frac{T}{k} \mathbf I,
    \quad
    \mathbf A^0
    =\frac{k}{T}
    \begin{pmatrix}
        D_0 & \kappa \\
        -\kappa & D_0
    \end{pmatrix}.
\end{equation}
The mean value during relaxation and the steady-state correlation are then written as
\begin{equation}\label{eq:ave_app}
    \langle z(\boldsymbol{x}(t))\rangle^\epsilon
    =\int d\boldsymbol{x}_0\,
    \langle z(\boldsymbol{x}(t))|\boldsymbol{x}_0\rangle
    P^\epsilon(\boldsymbol{x}_0)
\end{equation}
and
\begin{equation}\label{eq:corr_app}
    C_{z,z}^\textrm{s}(t,0)
    =\int d\boldsymbol{x}_0\,
    \langle z(\boldsymbol{x}(t))|\boldsymbol{x}_0\rangle
    z(\boldsymbol{x}_0)P^\textrm{s}(\boldsymbol{x}_0),
\end{equation}
respectively.
Using Eq.~\eqref{eq:cond_ave_x0_app}, we find
\begin{equation}
    \langle z(\boldsymbol{x}(t))\rangle^\epsilon
    =A^\textrm{c}
    \sqrt{\frac{\det(\mathbf Q^\epsilon)}
    {\det(\mathbf Q^\epsilon+\mathbf Q^\textrm{c}(t))}}-1
\end{equation}
and
\begin{equation}
    C_{z,z}^\textrm{s}(t,0)
    = A^\textrm{c}A^z
    \sqrt{\frac{\det(\mathbf Q^\textrm{s})}
    {\det(\mathbf Q^\textrm{c}(t)+\mathbf Q^z+\mathbf Q^\textrm{s})}}
    -A^\textrm{c}
    \sqrt{\frac{\det (\mathbf Q^\textrm{s})}
    {\det(\mathbf Q^\textrm{c}(t)+\mathbf Q^\textrm{s})}}
    -A^z
    \sqrt{\frac{\det (\mathbf Q^\textrm{s})}
    {\det(\mathbf Q^z+\mathbf Q^\textrm{s})}}
    +1.
\end{equation}
This gives
\begin{equation}\label{eq:nlfdr_analytic_app}
    \langle z(\boldsymbol{x}(t))\rangle^\epsilon
    =C_{z,z}^\textrm{s}(t,0)
    =\frac{\epsilon^2\kappa^2}
    {e^{t/\tau}(kD_0+\epsilon\kappa)^2-
    \epsilon^2\kappa^2},
\end{equation}
where
\begin{equation}
    \tau=\frac{T}{2 kD_0}.
\end{equation}

\subsection{Analytical solution for generalized Green--Kubo relation}

For the generalized Green--Kubo relation example, we consider $f(\boldsymbol{x})$ and $g(\boldsymbol{x})$ chosen from $x_1^2/2$ and $x_2^2/2$.  From the definition, the generalized drift and diffusion coefficients are given by~\cite{pavliotis2015stochastic,park2024stochastic}
\begin{equation}
    V^f(\boldsymbol{x})=L^\dagger f(\boldsymbol{x})
\end{equation}
and
\begin{equation}
    D^{f,g}(\boldsymbol{x})
    =L^\dagger(f(\boldsymbol{x})g(\boldsymbol{x}))
    -f(\boldsymbol{x})L^\dagger g(\boldsymbol{x})
    -g(\boldsymbol{x})L^\dagger f(\boldsymbol{x}).
\end{equation}
For $f=x_i^2/2$ and $g=x_j^2/2$ this gives
\begin{equation}
    V_i(\boldsymbol{x})
    \equiv V^{x_i^2/2}(\boldsymbol{x})
    =[-\mathbf A^\epsilon \boldsymbol{x}]_i x_i + D_0
\end{equation}
and
\begin{equation}\label{eq:general_Diff}
    D_{i,j}(\boldsymbol{x})
    \equiv D^{x_i^2/2,x_j^2/2}(\boldsymbol{x})
    =2 D_0 x_ix_j \delta_{i,j}.
\end{equation}
Combining Eq.~\eqref{eq:general_Diff} with Eq.~\eqref{eq:linear_cov_app}, we obtain
\begin{equation}
    \langle D_{i,j} (\boldsymbol{x}) \rangle^\textrm{s}
    = \frac{2 D_0^2 T}{k D_0 + \epsilon \kappa} \delta_{i,j}
\end{equation}

Next, using Eqs.~\eqref{expr:propagator_app} and \eqref{expr:ss_dist_app}, we obtain
\begin{equation}
    \langle \boldsymbol{x}(t) \boldsymbol{x}^\textrm{T}(0) \rangle
    = \mathbf{E}^\epsilon(t) \mathbf{C}^\epsilon,
\end{equation}
with $\mathbf{E}^\epsilon(t) = e^{- \mathbf{A}^\epsilon t}$.
Applying Gaussian factorization then yields
\begin{equation}
    \langle x_i(t)x_j(t)x_k(0)x_l(0)\rangle
    =C_{i,j}^\epsilon C_{k,l}^\epsilon
    +[\mathbf E^\epsilon(t)\mathbf C^\epsilon]_{i,k}
     [\mathbf E^\epsilon(t)\mathbf C^\epsilon]_{j,l}
    +[\mathbf E^\epsilon(t)\mathbf C^\epsilon]_{i,l}
     [\mathbf E^\epsilon(t)\mathbf C^\epsilon]_{j,k}.
\end{equation}
Using this relation, the centered correlation function is
\begin{equation}
    C_{\delta V_i,\delta V_j}^\textrm{s}(t,0)
    =\ [\mathbf A^\epsilon \mathbf E^\epsilon(t)\mathbf C^\epsilon (\mathbf A^\epsilon)^\textrm{T}]_{i,j}
    [\mathbf E^\epsilon(t)\mathbf C^\epsilon]_{i,j}
    +[\mathbf A^\epsilon \mathbf E^\epsilon(t)\mathbf C^\epsilon]_{i,j}
    [\mathbf E^\epsilon(t)\mathbf C^\epsilon (\mathbf A^\epsilon)^\textrm{T}]_{i,j}.
\end{equation}
The diagonal component is then given by
\begin{equation}
    C_{\delta V_i,\delta V_i}^\textrm{s}(t,0)
    =e^{-t/\tau_\epsilon}D_0^2
    \left(1+
    \frac{(k^2+\epsilon^2)(D_0^2+\kappa^2)}
    {(kD_0+\epsilon\kappa)^2}\cos(\omega t)
    \right),
\end{equation}
and the off-diagonal component is
\begin{equation}
     C_{\delta V_i,\delta V_j}^\textrm{s}(t,0)
    =e^{-t/\tau_\epsilon}D_0^2
    \left(1-
    \frac{(k^2+\epsilon^2)(D_0^2+\kappa^2)}
    {(kD_0+\epsilon\kappa)^2}\cos(\omega t)
    \right),   
\end{equation}
for $i\neq j$, where
\begin{equation}
    \tau_\epsilon=\frac{T}{2(kD_0+\epsilon\kappa)},
    \qquad
    \omega=\frac{2(k\kappa-\epsilon D_0)}{T}.
\end{equation}
After integrating the correlation functions over time, we obtain
\begin{equation}
    \int_0^\infty dt\,
    C_{\delta V_i,\delta V_i}^\textrm{s}(t,0)
    =\frac{D_0^2T}{kD_0+\epsilon\kappa}
\end{equation}
and
\begin{equation}
    \int_0^\infty dt\,
    C_{\delta V_i,\delta V_j}^\textrm{s}(t,0)
    =0,
\end{equation}
for $i \neq j$.

\begin{acknowledgments}
\FundingText
\end{acknowledgments}

\bibliographystyle{apsrev4-2}
\bibliography{main}

\end{document}